\documentclass[aip,graphicx]{revtex4-1}
\usepackage{graphicx}
\usepackage{braket}
\usepackage{upgreek}
\usepackage{natbib}
\usepackage{xcolor}

\draft 

\begin{document}


\title{Methods for Preparing Quantum Gases of Lithium} 



\author{Randall G. Hulet}
\email{randy@rice.edu}

\author{Jason H. V. Nguyen}
\author{Ruwan Senaratne}
\affiliation{Department of Physics and Astronomy, Rice University, Houston, Texas 77005, USA}


\date{\today}

\begin{abstract}
Lithium is an important element in atomic quantum gas experiments because its interactions are highly tunable, due to broad Feshbach resonances and zero-crossings, and because it has two stable isotopes, $^6$Li, a fermion, and $^7$Li, a boson.  Although lithium has special value for these reasons, it also presents experimental challenges.  In this article, we review some of the methods that have been developed or adapted to confront these challenges, including beam and vapor sources, Zeeman slowers, sub-Doppler laser cooling, laser sources at 671 nm, and all-optical methods for trapping and cooling.  Additionally, we provide spectral diagrams of both $^6$Li and $^7$Li, and present plots of Feshbach resonances for both isotopes.
\end{abstract}


\maketitle 

\section{\label{sec:intro}INTRODUCTION}

Lithium is popularly used in experiments with ultracold atoms.  Although there are several reasons for this, perhaps the most compelling is that lithium is found in nature in two isotopic forms: $^6$Li, a fermion, and $^7$Li, a boson. Furthermore, the two-body interactions of either of the isotopes are widely tunable using magnetic Feshbach resonances \cite{Tiesinga1993a,Chin2010a}. The ability to continuously tune interactions over a wide range and with high precision has enabled new experimental capabilities for both the bosonic and fermionic isotopes of lithium.
The isotopes of lithium have played an important role in the development of the field of atomic quantum gases.  The bosonic isotope was among the first to be cooled to quantum degeneracy \cite{Bradley1995,Bradley1997a} and was later used for studies of matter-wave solitons \cite{Khaykovich2002,Strecker2002}. The fermionic isotope, $^6$Li, was the second atomic Fermi gas to be cooled to degeneracy \cite{Truscott2001,Schreck2001} after $^{40}$K \cite{DeMarco1999a}. Later, the broad Feshbach resonance in $^6$Li was exploited to observe a strongly interacting superfluid in its expansion dynamics \cite{OHara2002,Kinast2004}, and to realize the BEC-BCS crossover \cite{Bartenstein2004a,Zwierlein2004} at about the same time as in $^{40}$K \cite{Regal2004}.

Although the lithium isotopes have much to offer, lithium presents a unique set of challenges that make it a relatively difficult atom to apply the standard methods of cooling and trapping.  Lithium has a relatively low vapor pressure, in comparison to the other alkali metals, and this necessitates high temperatures to produce sufficiently intense atomic beams or to create useful vapor cells. Consequently, care must be taken in the choice of construction and vacuum materials. The wavelength of its principal ($2S-2P$) transition, at 671 nm, is relatively short compared with all other alkali metals, with the exception of sodium, and consequently, there are relatively few laser sources available and the ones operating at the correct wavelength are less robust and powerful than those working in the more typical near-infra-red regime.  Lithium's light mass presents additional problems, including a large recoil energy that causes transverse spreading of a laser cooled atomic beam and the need for higher power lasers to produce an optical lattice with sufficient depth.  Finally, the hyperfine structure of lithium is anomalously small, preventing the straight-forward application of sub-Doppler laser cooling methods that are so important for cooling the other alkali metals.

In Secs. II--VII of this article, we will review the methods that we and others have developed to manage these obstacles. Our goal is to provide a compilation of the techniques that differ from the standard methods appropriate for most alkali species, but have proven to be the most effective for lithium.

\section{\label{sec:spect}SPECTROSCOPIC DATA AND COLLISIONAL PROPERTIES}

\subsection{\label{sub:spect}Spectroscopic data}

\begin{figure}
\includegraphics[width=\linewidth]{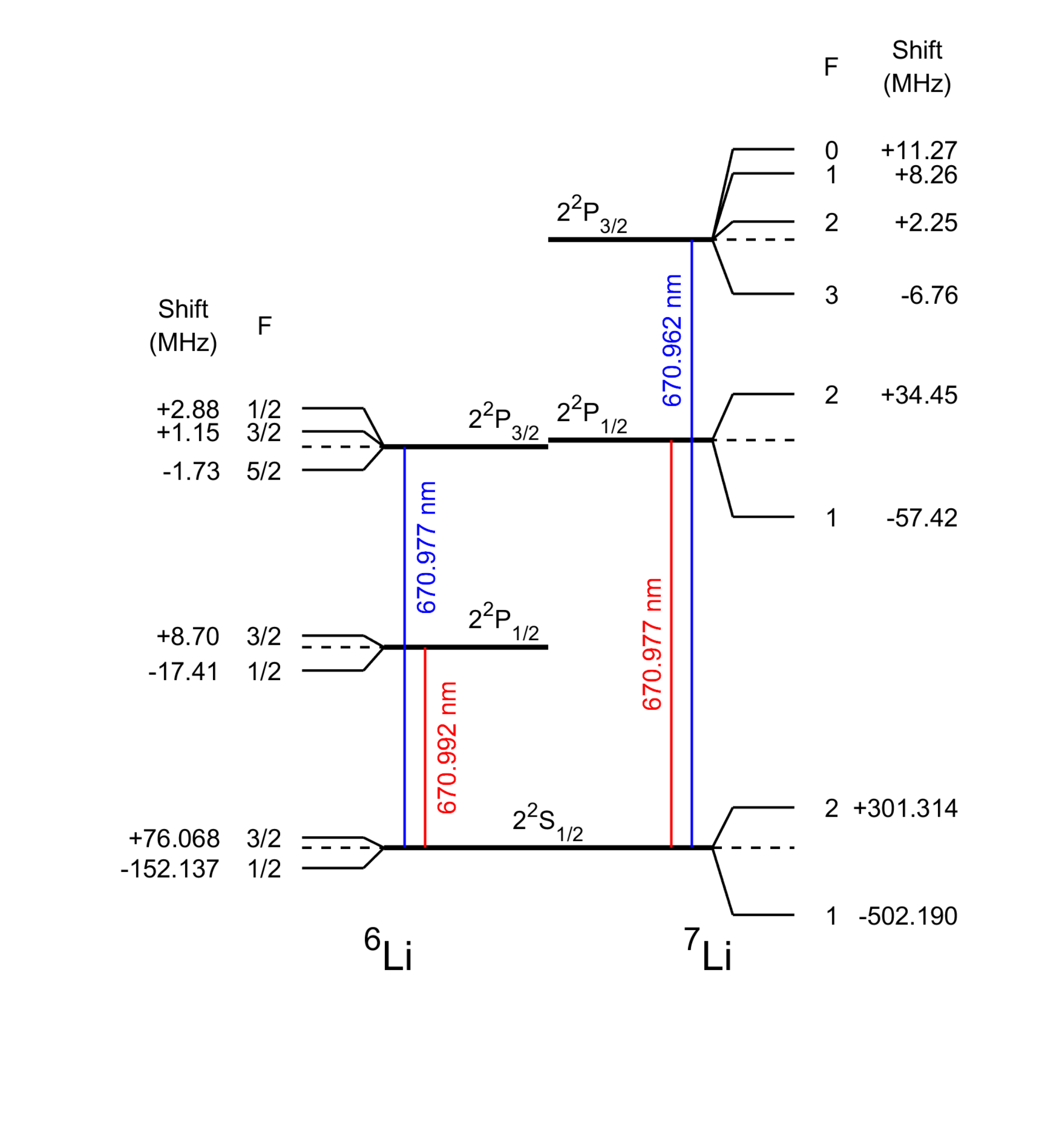}
\caption{Energy level structure of the low-lying states of $^6$Li and $^7$Li. The spectroscopic data is from Refs.~\citenum{Sansonetti2011} and \citenum{Beckmann1974}.  Blue: vacuum wavelengths of the D$2$ lines; red: vacuum wavelengths of the D$1$ lines. The 2P fine structure splittings are 10053 MHz for both isotopes.  The isotope shifts are 10534 MHz for both the D1 and D2 lines.}
\label{fig:levels}
\end{figure}

We begin by presenting the structure of the low-lying energy levels of lithium.  Figure \ref{fig:levels} gives the hyperfine structure of the $2S_{1/2}$ ground state, and the $2P_{1/2}$ and $2P_{3/2}$ excited states for both $^6$Li and $^7$Li.  The values of the various transition, hyperfine interval and isotope shift frequencies were measured using an optical frequency comb in a first-order Doppler-corrected atomic beam experiment \cite{Sansonetti2011}, except for the $2S_{1/2}$ hyperfine intervals which are from Ref.~\citenum{Beckmann1974}.  $^6$Li has a nuclear spin of $I=1$, resulting in a total angular momentum of either $F=1/2$ or $F=3/2$ in the $2S_{1/2}$ ground state, while $^7$Li has $I=3/2$ giving a ground state with either $F=1$ or $F=2$.


\begin{figure}
\includegraphics[width=\linewidth]{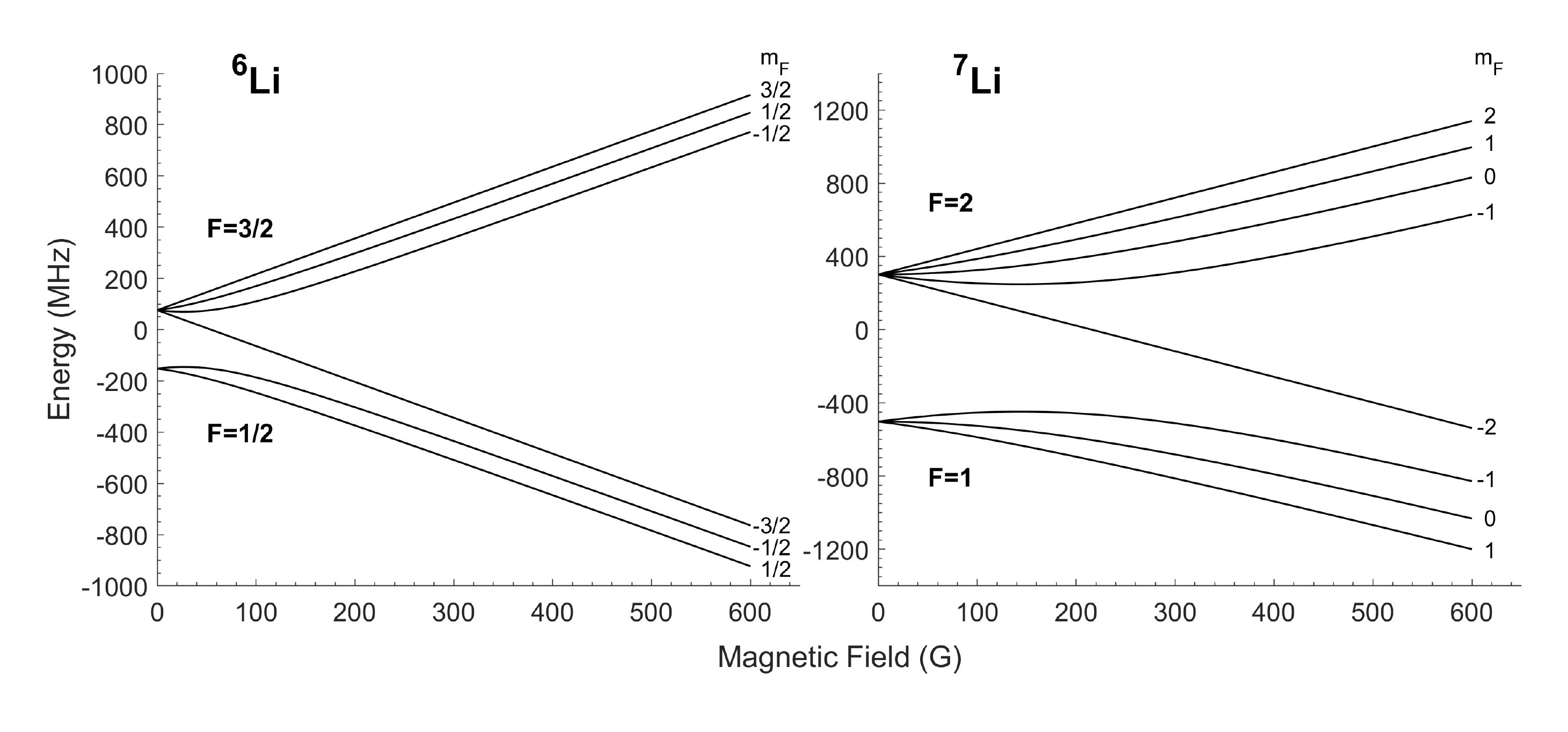}
\caption{Ground-state hyperfine sublevels of (a) $^6$Li and (b) $^7$Li in an applied magnetic field. The electron g-factor for both isotopes of lithium is $g_J = -2.0023010$, and the nuclear g-factors are: g$_I = 0.8219610$ for $^6$Li and g$_I = 2.1707235$ for $^7$Li \cite{Beckmann1974}.  We define the nuclear moment as $\mu_I = g_I \mu_N I$, where $\mu_N$ is the nuclear magneton. The ground state hyperfine splittings are 228.2052590(30) MHz for $^6$LI and 803.5040866(10) MHz for $^7$Li \cite{Beckmann1974}. }
\label{fig:hf}
\end{figure}

The ground-state hyperfine structure as a function of an applied magnetic field is shown for both isotopes in Fig.~\ref{fig:hf}. The spin projections $m_F$ exhibit a Zeeman structure that lifts the zero-field degeneracy of the ground state as described by the Breit-Rabi formula \cite{Ramsey1985}.

\subsection{\label{sub:scat}Scattering and interactions}

Two-body interactions and scattering are determined by an interaction potential $V(R)$.  The low energy scattering properties may be approximated by the $s$-wave phase shift, or more commonly, by the $s$-wave scattering length $a$.   In the case of alkali metal atoms, there are actually two ground-state potentials, $V_0$ and $V_1$, corresponding to either an electronic spin singlet state with $S=0$ or a spin triplet with $S=1$, and each has a corresponding scattering length.  Model potentials $V_0$ and $V_1$ for $^6$Li and $^7$Li were constructed using data mainly obtained from photoassociation measurements of lithium confined to a magneto-optical trap (MOT) \cite{Abraham1995,Abraham1995a,Abraham1997} and from the measured locations of Feshbach resonances and zero crossings~\cite{O'Hara2002a,Pollack2009,Gross2011,Dyke2013}.  The scattering lengths were extracted from these model potentials.

\begin{table}
  \begin{center}
    \caption{Singlet and triplet scattering lengths in units of the Bohr radius for isotopically pure and mixed gases of lithium~\cite{Abraham1997}.}
    \label{tab:table1}
    \begin{tabular}{l c c c}
		\hline\hline
      & $^6\mathrm{Li}$ & $^7\mathrm{Li}$ & $^6\mathrm{Li}$/$^7\mathrm{Li}$\\
      \hline
      $a_{1}$ & $-2160 \pm 250 $ & $-27.6\pm 0.5$ & $40.9\pm 0.2$ \\
      $a_{0}$  & $45.5 \pm 2.5 $ & $33\pm 2$ & $-20\pm 10$ \\
    \end{tabular}
  \end{center}
	\label{tab:scatlengths}
\end{table}

The triplet scattering lengths for $^6$Li and $^7$Li are both notable, but for different reasons.  The triplet scattering length for $^6$Li is $a_1$ = -2160(250) $a_{0}$ \cite{Abraham1997}, where $a_0$ is the Bohr radius.  Its extremely large magnitude indicates that a bound, or nearly bound state lies near the dissociation limit.  In this case, since $a_1 < 0 $, the molecular state lies just above the dissociation threshold. If $V_1$ were just 0.08 cm$^{-1}$ deeper, the virtual state would become bound and $a_1$ would be large and positive \cite{Abraham1997}. The triplet scattering length for $^7$Li is also negative, $a_1$ = -27.6(5) $a_{0}$, but relatively small in magnitude.  The fact that $a_1 < 0$ profoundly effects the nature of Bose-Einstein condensation for atoms interacting via the triplet potential, as it imposes a limit on the number of atoms that may form a stable Bose-Einstein condensate in a trapped gas \cite{Ruprecht1995,Bradley1997a,Perez-Garcia1998,Gammal2001,Parker2007}.

\begin{figure}
\includegraphics[width=\linewidth]{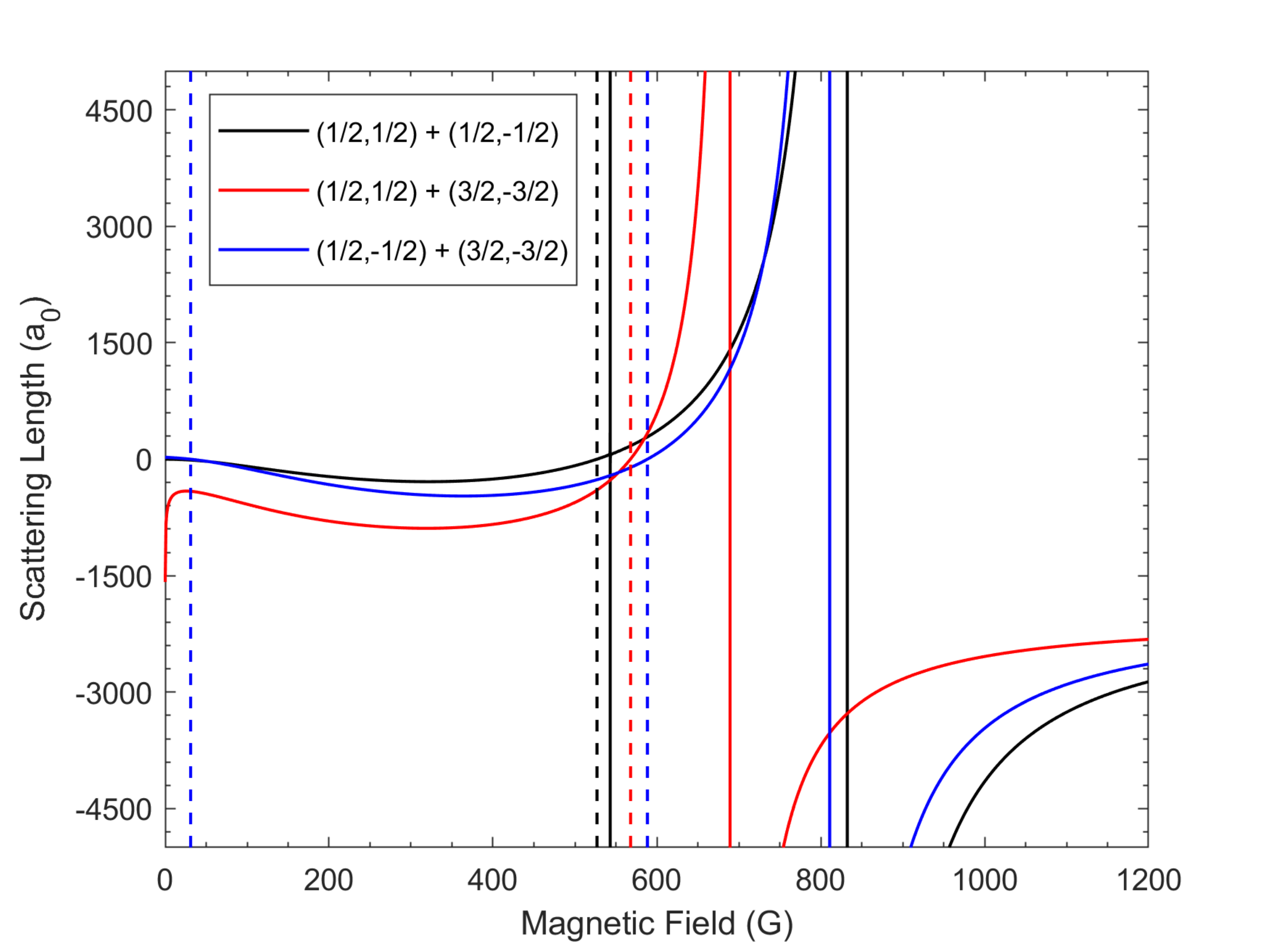}
\caption{S-wave Feshbach resonances involving the lowest three hyperfine levels of $^6$Li \cite{Houbiers1998}.  The levels are designated by the quantum numbers $(F,m_F)$. Note the narrow Feshbach resonance near $543\ $G for the $(1 /2, 1/2)+(1/2,-1/2 )$ pair.  Dashed vertical lines show positions of zero-crossings for each scattering length. These were calculated using the coupled-channels method \cite{Tiesinga1993a} with model potentials constructed from \textit{ab initio} calculations, various spectroscopic data, and measured locations of the resonances\cite{Konowalow1979,Schmidt-Mink1985,Barakat1986,Cote1994,Abraham1995,Abraham1995a,Yan1996,Abraham1997,Linton1999,Halls2001,Colavecchia2003,Pollack2009,Gross2011,Dyke2013,O'Hara2002a}.
 }
\label{fig:6FR}
\end{figure}

\begin{figure}
\includegraphics[width=\linewidth]{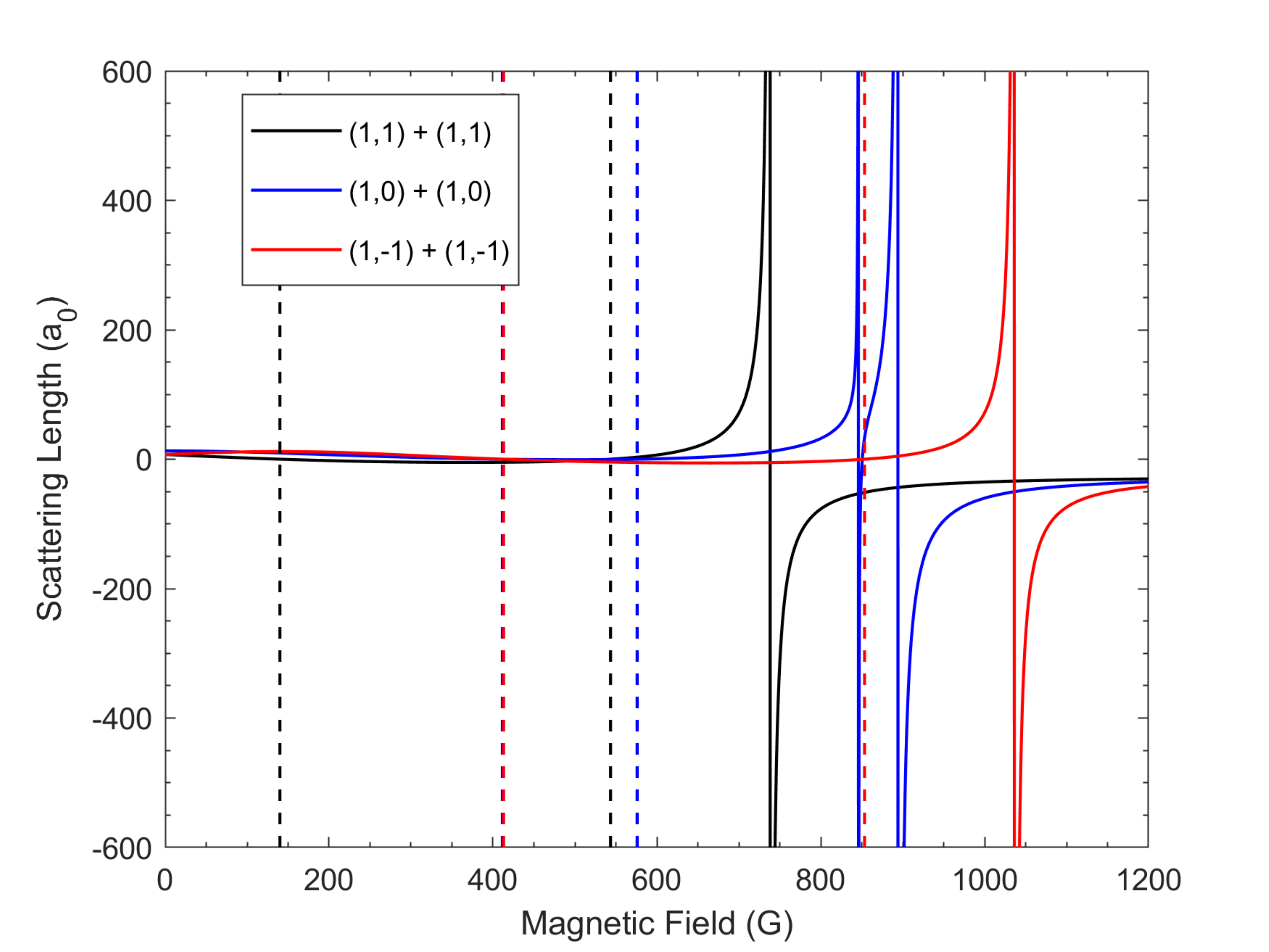}
\caption{Feshbach resonances for the lowest three hyperfine sublevels of $^7$Li for collisions between identical atoms. Otherwise same as Fig.~\ref{fig:6FR}.
}
\label{fig:7FR1}
\end{figure}

\begin{figure}
\includegraphics[width=\linewidth]{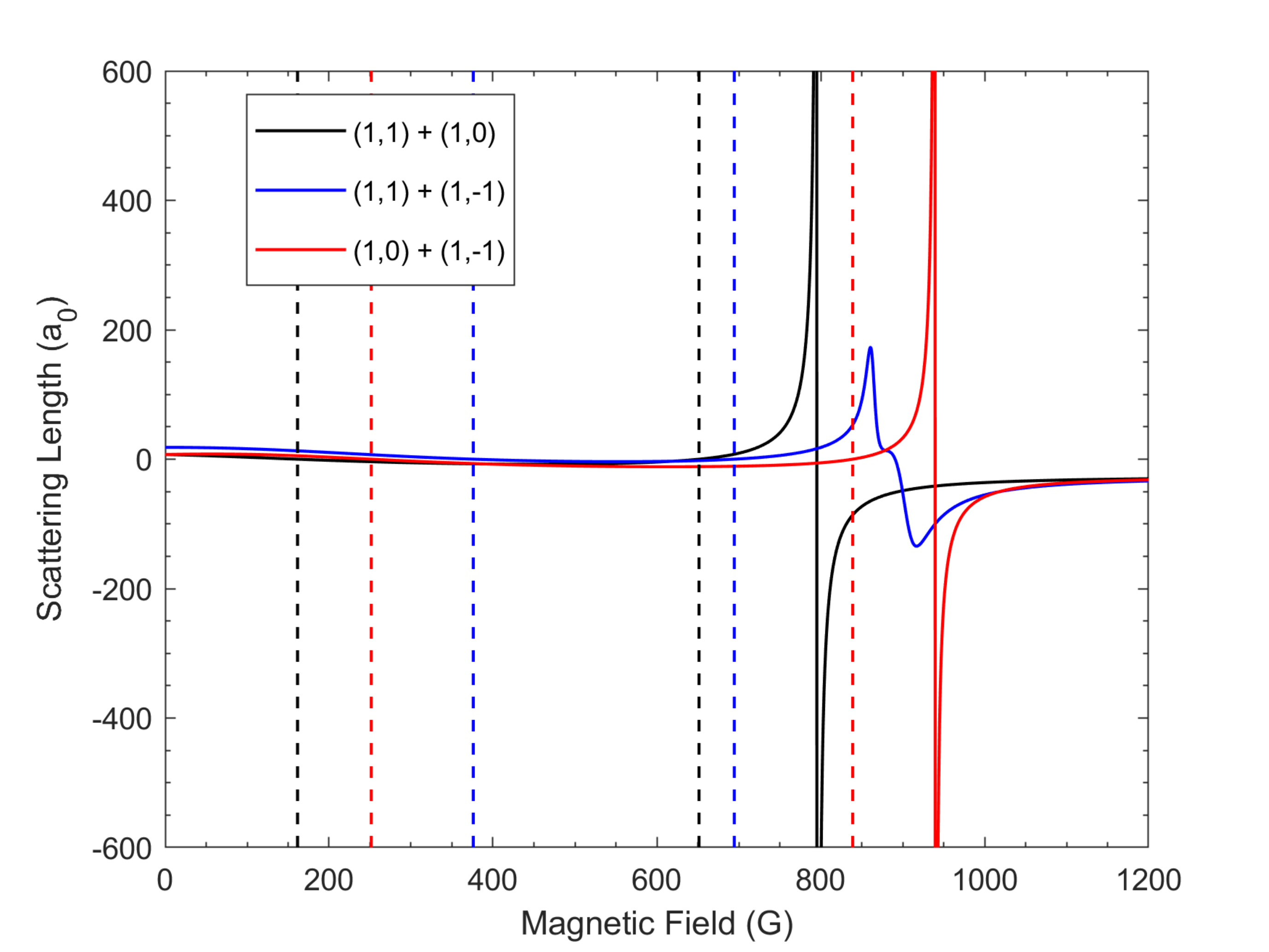}
\caption{Feshbach resonances for mixtures of the lowest three hyperfine sublevels of $^7$Li for collisions between atoms with differing spin. Otherwise same as Fig.~\ref{fig:6FR}.
}
\label{fig:7FR2}
\end{figure}

$S$ is only an approximate quantum number in the alkali metal atoms, as the two potentials $V_0$ and $V_1$ are weakly coupled by the hyperfine interaction \cite{Tiesinga1993a}. While the electron and nuclear spin aligned states, known as the stretched states, interact solely via $V_1$ none of the hyperfine sublevels interacts exclusively on $V_0$. The presence of two coupled interaction potentials, however, is extremely useful. A magnetic field may then be used to tune a bound state of the $V_0$ potential into resonance with the dissociation threshold of $V_1$ thus creating a tunable collisional ``Feshbach'' resonance \cite{Tiesinga1993a,Duine2004,Chin2010a}. Figure \ref{fig:6FR} shows the s-wave Feshbach resonances involving the three lowest hyperfine sublevels in $^6$Li \cite{Houbiers1998, Jochim2002, Bartenstein2005, Zurn2013}, while Figures ~\ref{fig:7FR1} and \ref{fig:7FR2}  show the Feshbach resonances for the lowest three sublevels of $^7$Li.

\section{\label{sec:beam}BEAM AND VAPOR SOURCES}

\begin{figure}
\includegraphics[width=\linewidth]{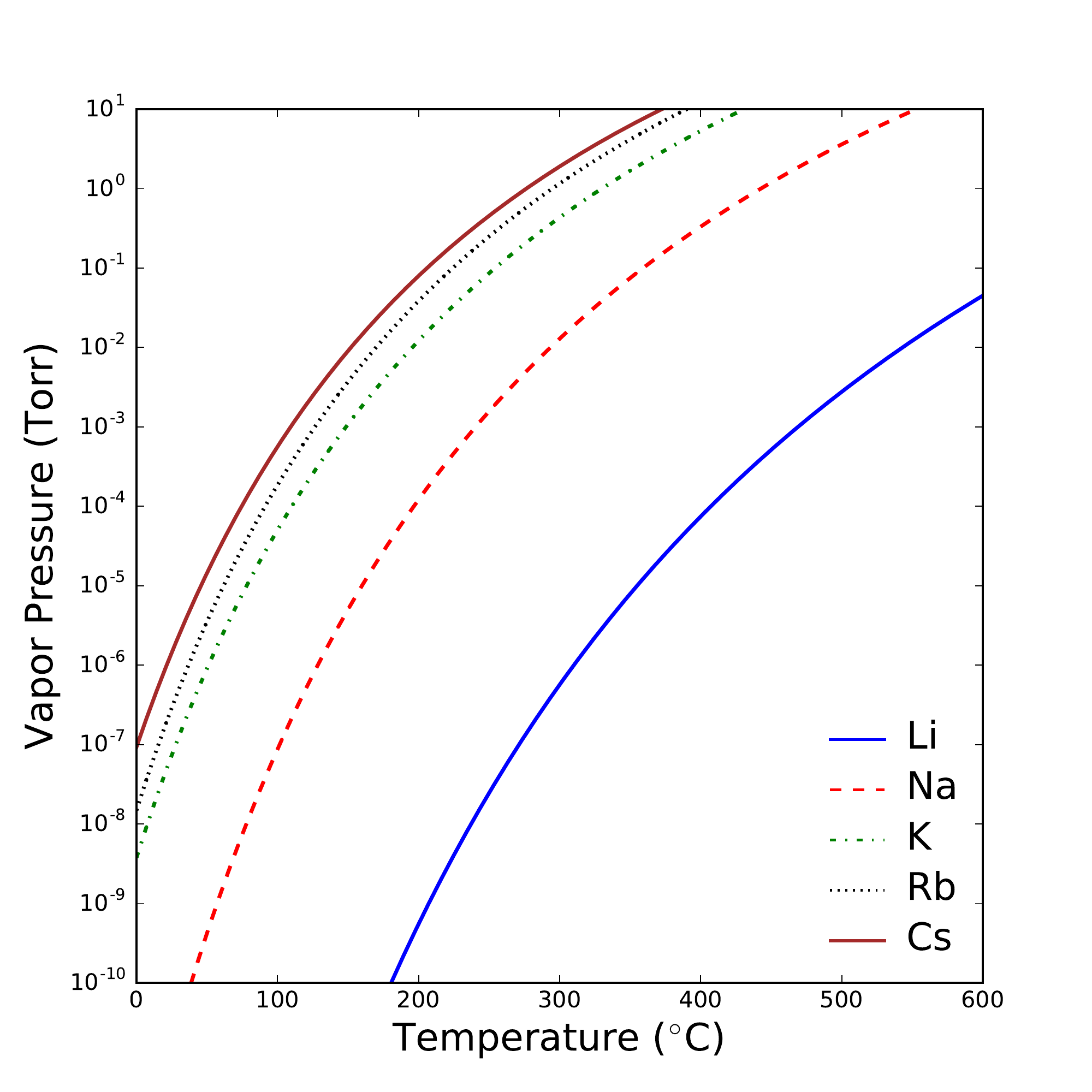}
\caption{Vapor pressure of the alkali metals from data tabulated in Ref.~\citenum{Nesmeyanov1963}.}
\label{fig:vapor}
\end{figure}

$^6$Li is distinctive because it is one of only two stable fermionic isotopes, along with $^{40}$K, among the alkali metals. Each has its own advantages and challenges.  An advantage for $^6$Li is that the natural abundance of $^6$Li is relatively high at 7.5\%, and furthermore, because of the usefulness of its high neutron absorption cross section to the nuclear industry, $^6$Li is readily available in an isotopically pure form. In comparison, $^{40}$K has a relative abundance of only $\sim$$10^{-4}$, but it can be obtained as a KCl salt that has been isotopically enriched to the level of 3-4.5\%.  The KCl salt may then be crafted into a dispenser of enriched $^{40}$K \cite{DeMarco1999b,Aubin2005}.  On the other hand, lithium requires a relatively high temperature of $\sim$600 C to produce a vapor pressure of 0.1 Torr, as shown in Fig.~\ref{fig:vapor}.

Lithium reacts with air, so it must be stored appropriately.  Typically, lithium is purchased in rod or wire form that is stored in mineral oil or packed in an argon environment.  Most of the mineral oil can be removed with petroleum ether while inside a glove bag purged with argon or other inert gas.  It takes 5-10 days of vacuum baking while heating the oven to typical operating temperatures to eliminate the oil contamination from the vacuum chamber.  We perform this bake using a gate valve to isolate the UHV portion of the chamber from the oven chamber as bake pressures will rise into the $\sim$$10^{-5}$ Torr range.  Lithium metal with natural isotope abundances ($92\%$ $^7$Li) can be purchased from ESPI Metals while isotopically pure $^6$Li can be obtained from Sigma Aldrich.

\subsection{\label{sub:oven}Atomic beam source}

\begin{figure}
\includegraphics[width=\linewidth]{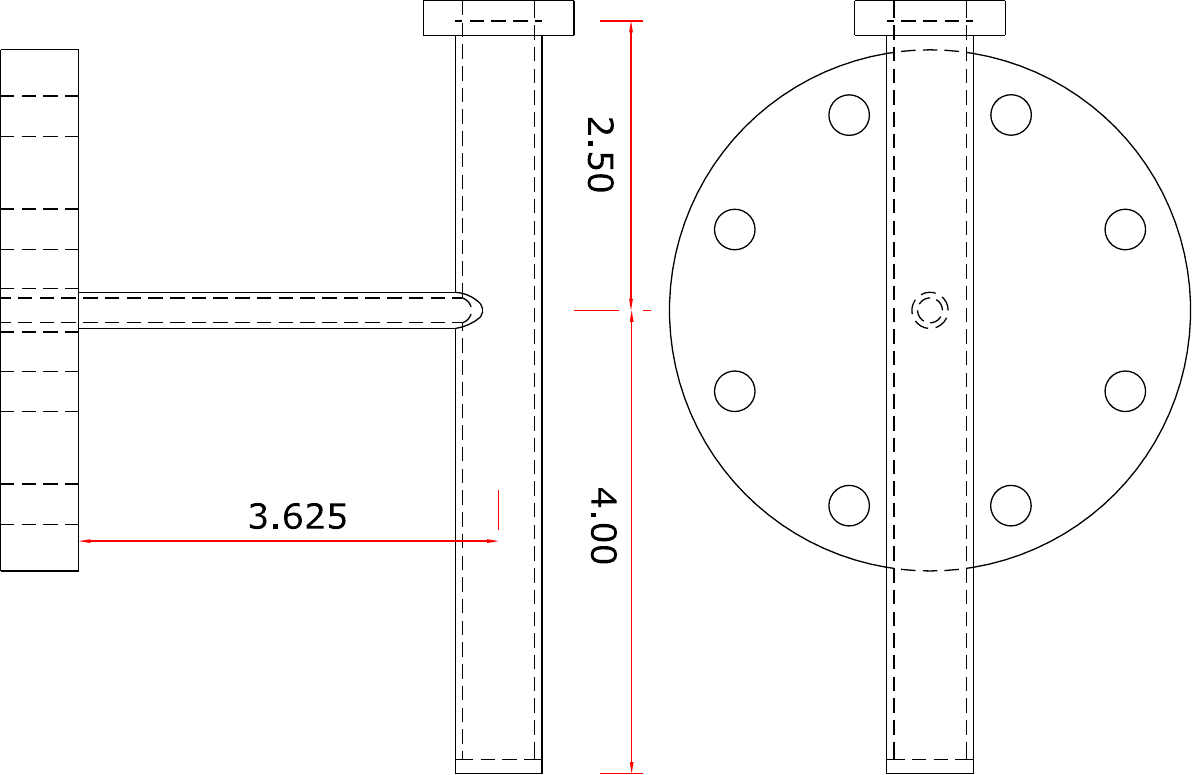}
\caption{Schematic drawing of the recirculating oven (all dimensions are in inches).  The $4.5$ in.\ rotatable flange attaches to the chamber.  The nozzle consists of stainless steel tube with an $OD=0.313$ in.\ and a wall that is $0.049$ in. thick.  The oven body has an $OD=0.750$ in. with a wall that is $0.060$ in. thick.  The top flange is a $1.33$ in. mini-flange.}
\label{fig:oven}
\end{figure}

Lithium's low vapor pressure influences the design of beam sources and vapor cells so that they are usable in the 500-600 C range. We have designed a simple recirculating oven for this purpose, as shown in the schematic drawing (Fig.~\ref{fig:oven}). Efficiency, simplicity, and compatibility with an ultra-high vacuum system (UHV) were the primary design considerations.  The oven consists of a reservoir for lithium made from an approximately $20$ mm diameter stainless steel tube and a smaller tube, functioning as a nozzle, that is welded at right angles into the middle of the reservoir.  The opposite end of the nozzle tube is welded into a through hole in the center of a UHV flange, which is joined to a stainless steel chamber, and sealed by a standard copper gasket/knife-edge assembly. Before installing the oven, it is also air-baked at $400\ $C for $36$ hours to reduce hydrogen outgassing during normal operation.  The lithium is added to the reservoir through a 1.33 in.\ UHV ``mini-flange" located at the top end end of the reservoir. As lithium is reactive in air, the reservoir is best filled in a glove bag purged with argon.

The output beam of the oven is designed to be well-collimated due to the small diameter of the nozzle output aperture compared to the length of the nozzle tube\cite{Ramsey1985}. A single layer of fine stainless steel mesh (SS304, \#80) is inserted into the nozzle tube to provide a return path for molten lithium to wick back to the reservoir. The mesh is extended only up to the outer surface of the mounting flange to prevent clogging at the cool end of the nozzle.  Wicking requires a temperature gradient along the nozzle tube. We use several heater tapes (Omega, model SST051) separately controlled by variable transformers to keep the central area of reservoir tube at $\sim$500 C, while the nozzle exit is kept at a lower temperature, but well above the melting point of lithium of 180 C (typically $\sim$300 C). The oven is covered by several layers of ceramic fiber insulation (McMaster-Carr 93315K34). The top 2-3 cm of the reservoir and the mini-flange are left exposed to air to prevent lithium from accumulating on the reservoir walls and potentially damaging the mini-flange seal. The temperature of the mini-flange is kept above 180 C. A layer of mesh is also placed around the inner wall of the reservoir up to 1 cm below the top of the mini-flange.  We cut a hole in this mesh at the location of the nozzle tube. Because of the relatively high oven temperatures, this flange is sealed with a nickel gasket, rather than a standard copper one.  This also minimizes the corrosion of the copper gasket from exposure to lithium vapor.

By operating the oven as described, lithium that is not within the small solid angle subtended by the oven nozzle exit is recirculated back to the oven, thus minimizing lithium consumption while producing a high central flux.  We have found that this oven will last for more than 5 years without service when loaded with $\sim$10 g of lithium. While the relatively large area of the nozzle helps to provide high flux, the aperture may be larger in diameter than the mean-free-path for elastic collisions, which would violate the effusive source criterion and alter the speed distribution \cite{Ramsey1985}.  Nonetheless, we find that we can load a MOT with $1.5 \times 10^9$ atoms in only 5 s \cite{Duarte2011} using a Zeeman slower as described in Sec.~\ref{sub:Zeeman}. At the time that this was written, Nor-Cal Products will construct such an oven from the drawings presented in Fig.~\ref{fig:oven} for \$400.  Alternative sources, such as one using multi-channel nozzles consisting of arrays of micro-capillaries have demonstrated excellent collimation of a lithium atomic beam while maintaining a similarly high flux\cite{Senaratne2015}.

\subsection{\label{sub:heatpipe}Lithium vapor cell}

Vapor cells in conjunction with saturated absorption spectroscopy are typically used as frequency references.  While simple all-glass cells are often used for this purpose with the other alkali metals, the temperature required to get sufficient optical depth with a lithium vapor generally prevents this simple solution.  In addition, lithium, like sodium, attacks glass and thus quickly renders windows unusable. A stainless steel tube, operating together with a buffer gas in the heat-pipe mode was found to be a trouble-free solution to these problems.

A simple design that we have successfully used consists of a stainless steel nipple 50 cm long with ISO KF-25 flanges attached at each end. Optical windows, mounted to KF-25 flanges, are attached to either end of the nipple to allow laser beams to pass.  The central $\sim$15 cm of this nipple is heated to $\sim$330 C to produce suitable absorption.  The windows are protected from lithium by flowing water through copper tubing soldered around an $\sim$8 cm length at each end of the nipple.  A buffer gas of $\sim$30 mTorr of argon provides a short mean-free path for lithium and thus prevents the lithium vapor from reaching the windows.

\section{\label{sec:Dopp}DOPPLER COOLING: PRINCIPAL TRANSITION}

Elements of the alkali metals were the first group of the periodic table to be laser cooled, trapped, and brought to quantum degeneracy (although trapped atomic ions were previously laser cooled).  Lithium beams were first transversely cooled \cite{Tollett1990b}, then slowed longitudinally, first using the Zeeman slowing method \cite{Lin1991}, and then followed by a chirp-cooling method \cite{Bradley1992}.

\subsection{\label{sub:redlasers}Laser sources at 671 nm}

The principal transition of lithium is the 2S-2P transition at 671 nm, as shown in Fig.~\ref{fig:levels}. The earliest laser cooling experiments with lithium used dye lasers pumped by an Ar ion laser \cite{Tollett1990b,Lin1991,Bradley1992}.  A dye laser could produce over 500 mW of red light. Shortly thereafter, however, extended cavity diode lasers (ECDL) were employed for detection using resonance fluorescence \cite{Bradley1990a,Chen1992,Bradley1992}. The performance of semiconductor lasers at this wavelength was significantly inferior to those operating at the principal transition wavelength of every other alkali atom, with the exception of sodium for which there is no direct band-gap materials available at 589 nm.  Maximum powers were limited to $\sim$10 mw, which is insufficient to slow an atomic beam or to make a robust magneto-optical trap.

Semiconductor tapered amplifiers were developed several years later \cite{Meyhus1992}.  These can amplify the weak signal from an ECDL to produce useable powers of $\sim$400 mW for near-infrared wavelengths \cite{Marquardt1996,Nyman2006} and $\sim$200 mW for the lithium wavelength \cite{Ferrari1999}. This master oscillator/power amplifier (MOPA) configuration was a satisfactory solution for more than a decade as Toptica Photonics and Eagleyard Photonics supplied tapered amplifiers that operate at 671 nm. Although supplied by both companies, the devices are apparently produced by Eagleyard.  Unfortunately, Eagleyard has experienced technical problems over the past several years, and has been unable to consistently produce a 671 nm TA device with the same quality and lifetime they previously attained.  As far as we know, Eagleyard is the sole source of TA's operating at red wavelengths. 

Two new developments partially mitigate this challenge.  First, M-Squared Laser offers a Ti-Sapphire laser whose tuning range can be extended down to 671 nm and secondly, Toptica has developed a system employing second harmonic generation of a MOPA operating near 1.34 $\mu$m.  Both manufacturers claim their systems can produce nearly 1 W at 671 nm. Another approach, commercially available from LEOS, is to begin with a 1342 nm ECDL, whose output is amplified by a Raman fiber amplifier, then doubled in a resonant cavity.   Such a system has demonstrated 2.5 W at 671 nm, and has been employed to produce a quantum gas of $^6$Li using all-optical methods \cite{Deng2015}.

\subsection{\label{sub:Zeeman}Zeeman slower}

The Zeeman slower \cite{Phillips1982} was one of the first methods developed to slow an atomic beam, and is still a commonly used and powerful method for loading MOT's. The Zeeman slower employs a magnetic solenoid in which the longitudinal magnetic field either increases or decreases as a function of position along its axis. The atomic beam passes through this solenoid, while a counter-propagating near resonant laser beam produces photon absorption/spontaneous emission cycles that slow the atoms.  The changing magnetic field is designed to compensate the changing Doppler shift of the slowing atoms by the position-dependent Zeeman field, thus keeping the atoms near resonance during their progression along the solenoid.

The velocity scale of lithium is high because it is light, and because the temperatures required to produce an appreciable vapor are high. The most probable velocity in a beam is $v_{p} = \sqrt{3k_{B}T/m}$, where $T$ is the temperature of the beam source, $m$ is the atomic mass, and $k_{B}$ is the Boltzmann constant \cite{Ramsey1985}. An oven temperature of $T \simeq 800$ K, for example, gives a high beam flux and at this temperature, $v_{p} \simeq 1800$ m/s and 1700 m/s for $^6$Li and $^7$Li, respectively.

The primary consequence of high beam velocities is the correspondingly large length needed for the Zeeman slower solenoid, $L = v_{0}^2/(2a)$, where $v_{0}$ is the initial velocity, or capture range of the slower, and $a = v_\textrm{rec}/(2\tau)$ is the maximum acceleration possible using the usual 2-level Doppler cooling \cite{Metcalf1999}.  Here, $v_\textrm{rec} = h/(m\lambda)$ is the recoil velocity of an atom from the absorption of a single resonant photon; for $^7$Li, $v_\textrm{rec} = 8.5$ cm/s, while for $^6$Li, $v_\textrm{rec} = 9.9$ cm/s.  Also, $\tau = 1/\gamma$ is the excited state lifetime, which for the lithium principal transition is 27.102(7) ns \cite{McAlexander1996}.

If we take $v_{0} = v_{p}$, we can capture a significant fraction of the atomic beam distribution, but the required length is long:  $L \simeq 90$ cm for both isotopes.  Not only would such a device occupy a large portion of an optical table, it would require significant electrical current and water cooling. Furthermore, as discussed below, transverse heating of the beam results in significant loss of beam intensity, an effect exacerbated by the long length.

Fortunately, it is not necessary to capture most of the atoms in the distribution to obtain a high flux of slow atoms.  Since the intensity distribution of an atomic beam source scales as $v^3$ for $v \ll v_{p}$  \cite{Ramsey1985}, the total number of slowed atoms $N$ is the integral of the distribution from $v = 0$ to $v = v_{0}$, thus giving $N \propto v_{0}^4$. At the same time, the solid angle subtended by the Zeeman slower exit aperture diminishes as $1/L^2 \propto 1/v_{0}^4$.  Thus, the gain in atom flux obtained by increasing $L$ is exactly cancelled by the reduction in solid angle.  The optimal length can then be kept short, as long as $L$ is greater than the distance between the oven aperture and the entrance to the Zeeman slower.

\begin{figure}
\includegraphics[width=\textwidth]{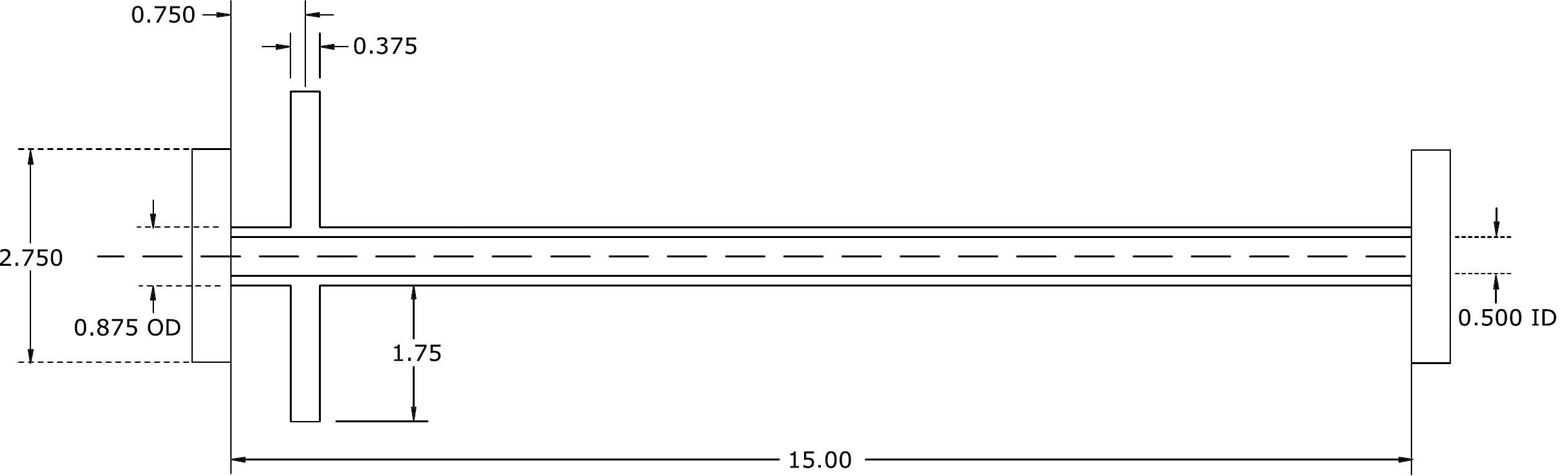}
\caption{Schematic drawing of the Zeeman slower showing the double wall construction. Atoms enter from left. Water enters through a cooling port. Two bare copper wires (No. 15 American wire gauge) are helically wrapped around and brazed to the outer surface of the inner tube in order to channel water down the slower and back to the outlet cooling port. All dimensions are in inches.}
\label{fig:slower}
\end{figure}

Our Zeeman slower design is shown in Fig.~\ref{fig:slower}.  It consists of a double-jacketed stainless steel vacuum nipple with 2 3/4 in. UHV flanges welded to either end.  Water flows through the inner jacket for cooling.  Magnet wire is wrapped around this nipple to generate an appropriately increasing field in the $\sigma^-$ configuration \cite{Barrett1991}. The magnet wire is installed using a thermally conductive potting compound.  The $\sigma^-$ configuration is better able to extract a slow, monoenergetic beam since the exit field falls rapidly from its peak value as atoms leave the slower causing their effective detuning with the slowing light to grow rapidly.  Thus, the atoms are quickly decoupled from the slowing light so they do not get turned around before arriving at the MOT.

Transverse heating produced by the laser photon absorption/spontaneous emission cycles poses a significant problem for lithium due to its relatively large $v_\textrm{rec}$ and $v_{0}$.  In this case, the transverse velocity, $v_{T}$, can grow to be comparable to the longitudinal exit velocity resulting in a significant loss of slowed atoms due to transverse spreading.  Because of the inherent randomness of the spontaneous emission process, $v_{T} = v_{rec} \sqrt{N_{ph}}$, where $N_{ph} = v_{0} / v_\textrm{rec}$ is the number of spontaneously scattered photons induced by the slowing laser.  Hence, $v_{T} = (v_\textrm{rec} v_{0})^{1/2}$.  Assuming $v_{0} \simeq 1000$ m/s gives $v_{T} \simeq 10$ m/s, which is comparable to a typical final longitudinal velocity. While we have been unable to eliminate this problem, it can be mitigated by using a 2D MOT located as close as possible to the exit of the Zeeman slower for beam collimation~\cite{Riss1990}.  In our systems, we incorporate a 2D MOT using a short vacuum nipple with 2 3/4 in. UHV flanges in which two pairs of small (1.3 in.) UHV flanges with optical viewports are mounted transverse to the atomic beam axis.

The lithium beam will coat the window that transmits the counter-propagating Zeeman slower laser beam, and without taking additional steps, the window eventually becomes opaque.  We minimize this problem in two ways. First, we mount the viewport at the end of a $\sim$1 m long UHV vacuum nipple to effectively reduce the solid angle subtended by the window, relative to the beam source.  Second, we use a sapphire vacuum viewport that is heated to $\sim$375 C to reduce the rate at which lithium adheres to the window.

\subsection{\label{sub:MOT}Magneto-optical trap}

The magneto-optical trap (MOT) \cite{Raab1987} is ubiquitous as it used in nearly every cold atom experiment. The MOT uses three pairs of counter-propagating laser beams in each of the three orthogonal directions. In this respect, the MOT resembles an optical molasses \cite{Chu1985} which provides velocity-dependent laser cooling in 3D. But in addition to laser cooling, the MOT uses an inhomogeneous magnetic field, produced by a pair of anti-Helmholtz coils, to create spatially-dependent radiative pressure to confine the atoms.

A standard lithium MOT, for which the numbers and temperatures are optimized by fixing the field gradient, laser intensities, and detunings, takes about 5 s to fully load from a Zeeman slower.  The maximum load is $\sim$$3 \times 10^{10}$ atoms, the temperature is typically 1-2 mK, and the peak density is $\sim$$8 \times 10^{10}$ cm$^{-3}$. However, by dynamically reducing the laser beam intensities and detunings, while increasing the magnetic field gradient, a compressed MOT with $10^{10}$ atoms, a temperature of $\sim$$700$ $\mu$K, and a peak density of $10^{11}$ cm$^{-3}$ can be attained \cite{Mewes1999,Truscott2001}.  As we will see in Sec. V, however, much higher phase space densities can be realized by using a gray or uv molasses.  Thus, these techniques can contribute to creating a much better starting point for evaporative cooling.

\section{\label{sec:sub-dopp}SUB-DOPPLER COOLING}

One of the most surprising observations in the early days of laser cooling was that the temperature of atoms in an optical molasses or a MOT could be much below the Doppler cooling limit, $T_{Dop} = {\hbar \gamma}/{2 k_B}$. For the alkali metals $T_{Dop} \simeq 100-250$ $\mu$K. However, careful measurements of the temperature of sodium released from an optical molasses found $T = 43 \pm 20$ $\mu$K, in strong disagreement with the Doppler limit of 240 $\mu$K for Na \cite{Lett1988}.

It was soon realized that this discrepancy is caused by the multi-level structure of the alkali metals arising from the hyperfine interaction and laser polarization gradients that can promote transitions between them \cite{Dalibard1989,Chu1998,CohenTannoudji1998,Phillips1998}.  While the Doppler cooling limit is predicated on an atom with a simple two-level structure with damping forces arising from a directionally-dependent Doppler shift, non-adiabatic optical pumping between hyperfine levels of an atom moving through a light field with polarization gradients can produce stronger cooling. In this case, temperatures can approach the recoil limit $k_{B} T_\textrm{rec} = \frac{1}{2} m v_\textrm{rec}^2 = h^2/(2m\lambda^2)$,  where $T_\textrm{rec}$ for the alkali metals varies from 3.5 $\mu$K for $^6$Li to 100 nK for Cs.  These temperatures are 40-1000 times lower than $T_{Dop}$.  The recoil limit can be approached with the appropriate configuration of laser polarizations, and with sufficient hyperfine splitting of the sublevels.  The excited state hyperfine structure in lithium is nearly unresolved rendering sub-Doppler cooling much less effective than in other alkali metals, and considerably more difficult to implement.  Nontheless, sub-Doppler cooling of lithium has recently been achieved \cite{Grier2013,Hamilton2014}.

\subsection{\label{sub:gray}Gray molasses}

Gray molasses, or $\Lambda$-enhanced sub-Doppler cooling, has been demonstrated as an effective cooling technique for Li. This technique requires a $\Lambda$-type three-level system, a requirement which is satisfied in the alkalis thanks to the ground-state hyperfine structure. For the electronically excited level it is convenient to select the  $2\,^2$P$_{1/2} \ket{F'=I+1/2}$ level, as it is well-resolved (the prime refers to the $2\,^2$P$_{1/2}$ state). Two laser beams address the $\ket{F=I+1/2}\rightarrow\ket{F'=I+1/2}$ and $\ket{F=I-1/2}\rightarrow\ket{F'=I+1/2}$ transitions. Coherent superpositions of the two ground states may form a bright or a dark state when the lasers are in resonance with the two-photon transition. The energy of the bright state is modulated by the laser intensities, while that of the dark state is not. By applying an appropriately blue-detuned bichromatic lattice consisting of these two frequencies, it is possible to remove energy from the atoms by transferring them from the bright state to the dark state at regions of high intensity. Motional coupling at regions of low intensity can cause dark-state atoms to be transferred to the bright state. This allows for Sisyphus-like cooling, as well as velocity-selective coherent population trapping (VSCPT) to occur\cite{Aspect1988,Weidemuller1994}.


Temperatures of $\sim$$60\,\upmu$K and phase-space densities of $\sim$$10^{-5}$ have been achieved with gray molasses for both isotopes of Li\cite{Grier2013,Burchianti2014,Satter2018,Long2018}. The density of these clouds is limited by radiation trapping to $\sim$$10^{10}$ cm$^{-3}$. In order to transfer a significant number of atoms into an optical trap requires a large trapping volume and, therefore, high-power trapping beams. With the availability of 200 W fiber lasers near 1070 nm, gray molasses has become a viable path to eliminating the need for an intermediate stage of magnetic trapping. The ac Stark shifts induced by the optical trapping beams of the one-photon transitions have been reported in Ref. \citenum{Burchianti2014} to be +6.3(7) MHz/(MW/cm$^2$) at 1073 nm, and can be effectively counteracted by a relatively modest frequency shift of the gray molasses beams as the optical trap is ramped up. Using this technique, the authors report an optical trap containing $2 \times 10^7$ $^6$Li atoms at $80\,\upmu$K, and following evaporation a degenerate Fermi gas consisting of $7 \times 10^5$ atoms.

\subsection{\label{sub:narrow}Narrow-line cooling on the 2S-3P transition}

Lower Doppler limited temperatures may be realized by using a narrower transition than the usual principal transition.  This approach was recently exploited in $^{40}$K using the 4S-5P transition \cite{McKay2011a} and in $^6$Li using the 2S-3P transition \cite{Duarte2011} to achieve lower temperatures as a final stage of magneto-optical trapping.  Although these transitions are still dipole allowed, the dipole matrix element between the nS ground state and the (n+1)P excited state is significantly weaker than for the nS-nP principal transition. In the case of lithium, the 3P excited state has a natural linewidth of 754 kHz, which is $\sim$8 times narrower than the 2P state \cite{Kramida2018}.  The corresponding Doppler limit of $T_{Dop}\simeq 18$ $\mu$K is consequently 8 times lower than for a 2S-2P ``standard" MOT.

The 2S-3P transition wavelength of 323 nm is too far into the UV for fundamental laser sources, but $\sim$10-100 mW at 323 nm can be generated by frequency doubling the output of a 646 nm laser source that is generated either by an ECDL MOPA operating in the red, or by frequency doubling a $1.3$ $\mu$m laser.  While a UV laser source is somewhat inconvenient and expensive, the shorter wavelength results in a smaller absorption cross section which enables laser cooling a lithium vapor to higher densities, and therefore, to higher phase space densities.  A UV MOT was demonstrated with a density of $2.9 \times 10^{10}$ cm$^{-3}$, a temperature of 59 $\mu$K, and a corresponding phase space density $\rho_{ps} = 2.3 \times 10^{-5}$.  For comparison, the phase space density for a compressed red MOT in the same apparatus was approximately 10 times less \cite{Duarte2011}.

\subsection{\label{sub:magic}Magic wavelength}

Perhaps even more significant for laser cooling on the 2S-3P transition is the existence of a ``magic wavelength" where the differential ac Stark shift between the upper and lower states vanish \cite{Ido2000,Grain2007}. A magic wavelength for the 2S$_{1/2}$-3P$_{3/2}$ transition was predicted at 1071 nm \cite{Safronova2012}, as was subsequently verified experimentally \cite{Duarte2011}.  By optically trapping lithium atoms with a laser operating at a 1071 nm wavelength, atoms may be continuously cooled on the UV transition as they load the trap.  With this scheme, a quantum degenerate gas with $3 \times 10^6$ $^6$Li atoms was produced in 11 s by evaporating in a crossed-beam optical trap operating at the magic wavelength \cite{Duarte2011}.

\section{\label{sec:trap}TRAPPING AND EVAPORATIVE COOLING}

While the MOT is a general purpose trap with many applications, it is not capable of achieving sufficiently high phase space density to produce quantum degeneracy because the optical density becomes $\sim 1$ for $n \geq 3\times 10^{10} \ \mathrm{cm^{-3}}$.  All experiments creating ultracold atomic quantum gases use evaporative cooling in conjunction with either a magnetic or a pure optical trap that is often loaded by a MOT \cite{Hess1986, Ketterle1996, Sackett1997}.

\subsection{\label{sub:magtrap}Magnetic traps}

The first quantum gas of lithium was made in a magnetic trap constructed with six permanent magnets \cite{Tollett1995}.  The trapping geometry was that of an Ioffe-Pritchard trap, which features a potential with quadratic curvature combined with a uniform bias field \cite{Pritchard1983}.  Because of the strength of permanent magnets, the trapping potential had both a large depth and  volume.  It was loaded directly from a slowed atomic beam of $^7$Li, and laser cooled to a temperature of $\sim$1 mK without a MOT stage.  The only magnetically trappable hyperfine sublevel of $^7$Li stable to spin exchange collisions is the ($F=2, m_F=2)$ state, denoted as (2,2), although the (2,2) state can undergo loss-producing two-body dipolar decay collisions \cite{Gerton1999}. The rate of dipolar decay combined with the small $s$-wave triplet scattering length of only $-27.6\ a_{0}$ prevents $^7$Li from undergoing runaway evaporative cooling for which the elastic scattering rate exceeds the collisional loss rate leading to increasing density as evaporation proceeds \cite{Sackett1997}. While Bose-Einstein condensates could be produced in a permanent magnet trap \cite{Bradley1995, Bradley1997a}, its inability to be shut off eliminated time-of-flight as a tool to measure the momentum distribution, and prevented the transfer of the atoms to an optical trap where the field may be tuned to a Feshbach resonance.

Fermi-Dirac statistics prevent $s$-wave interaction between identical fermions.  In order to evaporatively cool a Fermi gas, one must employ either a two spin-state mixture \cite{DeMarco1999a}, or sympathetic cooling of the fermions using a Bose gas.  $^6$Li in the ($3/2, 3/2$) state was sympathetically cooled by $^7$Li in the ($2,2$) in a magnetic trap \cite{Truscott2001,Schreck2001b}. The interspecies scattering length is shown in Table \ref{tab:table1} to be 41 $a_{0}$, which is sufficient to perform efficient sympathetic cooling.

\subsection{\label{sub:ODT}Optical dipole traps}

While magnetic traps provide a path to quantum degeneracy, in many experiments, atoms are transferred following evaporation from the magnetic trap to an optical dipole trap where a tunable external magnetic field may be applied without effecting the trap.  This is usually the case in experiments involving Feshbach resonances, for example. The optical dipole trap is a conservative potential that arises from mixing of the ground and excited states by a far-detuned laser \cite{Dalibard1985, Chu1986}.  The potential depth of an optical dipole trap scales as $I/\Delta$, where $I$ is the laser intensity, and $\Delta$ is the detuning of the laser relative to resonance of an effective two-level system; $\Delta$ is made sufficiently large to minimize spontaneous emission, which scales as $1/\Delta^2$.  For a focused red-detuned laser beam, atoms are attracted to the intensity maximum, while being repelled by a blue-detuned beam. Unfortunately, the potential depth of an optical dipole trap is typically much smaller than for a magnetic trap.  Using lithium as an example, a laser beam with 1 W of power, focused to a Gaussian radius of 50 $\mu$m and a wavelength of $1064$ nm produces a potential depth of only $\sim$15 $\mu$K.  While it is simpler to transfer directly from a MOT to an optical dipole trap, thus realizing an all-optical system, the depth of the optical trap is often insufficient to contain the thermal energy distribution from the MOT.  As described in Sections~\ref{sub:gray} and \ref{sub:narrow}, either a gray molasses cooling scheme or narrow-line Doppler cooling on the 2S-3P transition now provide an all-optical pathway to quantum degeneracy.  The first all-optical lithium experiment used a high-power $\mathrm{CO_2}$ laser to create a deep optical potential \cite{Ohara1999,Granade2002}.

\section{\label{sec:PPCI}POLARIZATION PHASE CONTRAST IMAGING}

Optical absorption or phase-contrast imaging of the \textit{in-situ} density or the momentum distribution of the atoms in time-of-flight provides valuable information.  Generally, a near-resonant probe laser passing through the atoms is attenuated and it acquires a phase shift, both of which may be exploited to extract information about the atomic sample.  To account for both effects, we introduce a complex phase, $\beta = \phi + i\alpha / 2$, where $\phi$ is the dispersive phase shift and $\alpha$ is the absorption coefficient resulting from spontaneous emission.  The transmitted field is thus described as $\vec{E}=\vec{E_0} e^{i\beta}$.  Imaging the atoms onto a CCD camera produces an absorption signal, $I_{S}=|\vec{E}|^2=I_0 e^{-\alpha}$, where $I_0=|E_0|^2$.  The absorption signal is independent of the acquired phase, and scales with the probe detuning from resonance $\Delta$ as $\Delta^{-2}$.

Absorption imaging is commonly employed because of its simplicity: one simply images the shadow cast by the atoms.  It has two primary deficiencies, however.  Firstly, it is destructive since it relies on spontaneous emission to generate the absorption signal. Secondly, since $\phi$ falls off more slowly with detuning, as $\Delta^{-1}$, $\phi$ may be as large as ~$\pi/2$ or greater, especially for higher densities, and distortions will occur.  Distortions are lessened by reducing the density by time-of-flight expansion.  For \textit{in-situ} images, however, the density is often too large to have an ample absorption signal while simultaneously having sufficiently small dispersive distortion.  In these cases, it can be advantageous to employ phase-contrast imaging where the signal depends on $\phi$ in addition to, or instead of $\alpha$.  Furthermore, since $\phi$ does not depend on spontaneous emission, phase-contrast may be used to take multiple, minimally-destructive images.

In the usual implementation of phase-contrast imaging a small $1/4$-wave plate is placed at the focus of the probe beam after passing through the atom cloud, so that the phase of the unscattered light is shifted by $\pm \pi /2$ ~\cite{Andrews1997}.  This results in an interference between the scattered light and the probe field so that $I_{S}= I_0 e^{-2\phi}$ depends solely on $\phi$, rather than $\alpha$.  For large enough $\Delta$, $\alpha \ll |\phi| \ll 1$, and the signal is linear in $\phi$.

A more flexible phase-contrast method, polarization phase-contrast imaging, or PPCI~\cite{Bradley1997a, Sackett1997a}, exploits the birefringence of the atoms in the presence of a strong magnetic field and does not require a phase plate.  When the Zeeman shift is large compared to the excited state linewidth, $\gamma$, the atoms polarize according to their $m_F$ value.  The interaction between the atoms and the probe beam depends on the polarization of the probe field, which decomposes into two elliptical polarizations, one that couples to the transition dipole and scatters, and one that does not. The coupled component picks up a phase shift, while the uncoupled component serves as a reference field.  The two components are combined and interfered by passing them through a linear polarizer.  The angle of the polarizer with respect to the incident probe polarization determines the relative contribution to the detected signal to terms proportional to $\phi$, as in linear phase contrast imaging, and to $\phi^2$, as for dark-field imaging~\cite{Andrews1996}.  Thus, a simple adjustment of the polarizer angle controls the character of the image and is easily optimized.  This technique has also been referred to as Faraday imaging~\cite{Gajdacz2013, Kaminski2012}.

\section{\label{sec:conclusion}CONCLUSION}

In this article, we reviewed the methods and apparatus developed over the past 30 years to effectively cool, trap, and detect quantum gases of lithium.  Our goal is to collect a record of best practices to assist future experimenters to navigate this complex set of challenges.  While the approaches that we describe have been immensely successful, we expect that what we have written is not the last word, and that improvements will lead to even faster cooling and trapping cycles, more robust laser systems, and lower temperatures to access previously unexplored quantum states of matter.

\section*{\label{sec:acknowledgments}ACKNOWLEDGMENTS}
We are grateful to Eduardo Ibarra for help with the figures.  This work was partially supported by the NSF (Grant No. PHY-1707992), the Army Research Office Multidisciplinary University Research Initiative (Grant No. W911NF-14-1-0003), the Office of Naval Research, and The Welch Foundation (Grant  No. C-1133).

\end{document}